\input{aipcheck}
\documentclass[
    ,final            
  ]
  {aipproc}

\layoutstyle{8x11single}

\usepackage{amsmath, amsfonts, amssymb, graphicx}
\usepackage{bbold}

\newcommand\bi{\begin{itemize}}
\newcommand\ei{\end{itemize}}
\newcommand\bb[1]{\begin{block}{#1}}
\newcommand\eb{\end{block}}
\newcommand\be{\begin{equation}}
\newcommand\ee{\end{equation}}
\newcommand\bea{\begin{eqnarray}}
\newcommand\eea{\end{eqnarray}}

\newcommand\Ham{\hat{H}}

\newcommand\nr{{\it n}({\bf r})}
\newcommand\kv{{\bf k}}
\newcommand\rv{{\bf r}}

\newcommand\vev[1]{\left\langle #1\right\rangle}

\begin{document}

\title{Matrix Structure Exploitation in Generalized Eigenproblems Arising in Density Functional Theory}


\author{Edoardo Di Napoli}{
  address={RWTH Aachen, AICES, Schinkelstr. 2, 52062 Aachen, Germany}
}

\author{Paolo Bientinesi}{
  address={RWTH Aachen, AICES, Schinkelstr. 2, 52062 Aachen, Germany}
}

\begin{abstract}
In this short paper, the authors report a new computational approach in the context of Density Functional Theory (DFT). It is shown how it is possible to speed up the self-consistent cycle (iteration) characterizing one of the most well-known DFT implementations: FLAPW. Generating the Hamiltonian and overlap matrices and solving the associated generalized eigenproblems $Ax = \lambda Bx$ constitute the two most time-consuming fractions of each iteration. Two promising directions, implementing the new methodology, are presented that will ultimately improve the performance of the generalized eigensolver and save computational time. 
\end{abstract}

\maketitle


\section{Introduction to Density Functional Theory}

Density Functional Theory (DFT)~\cite{DG} is a powerful method of investigation that has become the ``standard model'' of material science. DFT is one of the most effective frameworks for studying complex quantum mechanical systems. DFT-based methods are growing as the standard tools for simulating new materials. The core of the method relies on the simultaneous solution of a set of Schr\"odinger-like equations. These equations are determined by a Hamiltonian operator $\Ham$ containing an effective potential $v_0[n]$ that depends functionally on the one-particle electron density $\nr$. In turn, the wave functions $\phi_i({\bf r})$, which solve for the equations, compute the one-particle electron density used in determining the effective potential. The latter depends explicitly on the external atomic Coulomb potential $v_I$ and the Hartree term $w({\bf r, r'})$ describing interactions between electrons.

\begin{eqnarray}
\begin{array}[l]{l}
		v_0([n],{\bf r}) = v_I({\bf r}) + \int w({\bf r, r'}) \nr + v_{xc}([n],{\bf r}) \\
		\vspace{0.1cm}
		\Ham \phi_i({\bf r}) = \left( -\frac{\hbar^2}{2m} \nabla^2 + v_0({\bf r}) \right) \phi_i({\bf r}) = \epsilon_i \phi_i({\bf r}) \quad ; \quad \epsilon_1 \leq \epsilon_2 \leq \dots\\
		\vspace{0.1cm}
		\nr = \sum_i^N |\phi_i({\bf r})|^2
\end{array}
\end{eqnarray}

In practice, this set of equations, also known as Kohn-Sham (KS)~\cite{KS}, is solved self-consistently; an educated guess for $\nr$ is used to calculate the effective potential $v_0$ that, in turn, is inserted in the Schr\"odinger-like equations whose solutions, $\phi_i({\bf r})$, are used to compute a new charge density $n'({\bf r})$. Convergence is checked by comparing the new density to the starting one. If convergence is not reached, an opportune mixing of the two densities is selected as a new guess, and the process is repeated. This is properly called an outer-iteration of the DFT self-consistent cycle.

In principle, the theory only requires as input the quantum numbers and the positions of the atoms constituting the investigated system. In practice, DFT implementations depend on the particular modeling of the atomic structure and the orbital basis used to parametrize it. The Full-potential Linearized Augmented Plane Wave (FLAPW) method~\cite{FKWW,FJ} is one of the many techniques implementing DFT that is based on plane wave expansion of $\phi_{\kv,\nu}(\rv)$, where the Bloch vector $\kv$ and the band index $\nu$ replace the generic index $i$. The Bloch wave function $\phi_{\kv,\nu}(\rv) = \sum_{|{\bf G + k}|\leq {\bf K}_{max}} c^{\bf G}_{\kv,\nu} \phi_{\bf G}(\kv,\rv)$ is expanded in terms of a basis set $\phi_{\bf G}(\kv,\rv)$ indexed by vectors ${\bf G}$ lying in the lattice reciprocal to configuration space. In FLAPW the configuration (physical) space of the quantum sample is divided into spherical regions, called Muffin-Tin (MT) spheres, centered around atomic nuclei, and interstitial areas between the MT spheres. The basis set $\phi_{\bf G}(\kv,\rv)$ takes a different expression depending on the region
\begin{eqnarray}
	\begin{array}[l]{lr}
	e^{i({\bf k+G})\rv} & \qquad \textrm{Interstitial}\\
	\displaystyle\sum_{\it l,m} \left[a^{\alpha,{\bf G}}_{\it lm}(\kv) u^{\alpha}_{\it l}(r) 
	+ b^{\alpha,{\bf G}}_{\it lm}(\kv) \dot{u}^{\alpha}_{\it l}(r) \right] Y_{\it lm}(\hat{\bf r}_{\alpha}) & \qquad \textrm{Muffin Tin}\\
	\end{array}
\end{eqnarray} 
where the coefficents $a^{\alpha,{\bf G}}_{\it lm}(\kv)$ and $b^{\alpha,{\bf G}}_{\it lm}(\kv)$ are determined by imposing continuity of $\phi_{\bf G}(\kv,\rv)$ and its derivative at the boundary of MT. Due to this expansion the KS equations naturally translate to a set of generalized eigenvalue problems $\sum_{\bf G'} \left[ A_{\bf GG'}(\kv) - \lambda_{\kv\nu} B_{\bf GG'}(\kv) \right] c^{\bf G'}_{\kv\nu} =0$ for the coefficients of the expansion $c^{\bf G'}_{\kv\nu}$ where the Hamiltonian and overlap matrices $A$ and $B$ are given by multiple integrals and sums
\be
	\{A(\kv),B(\kv)\} = \sum \iint   \phi^{\ast}_{\bf G}(\kv,\rv) \{\Ham,\hat{\mathbb{1}}\}  \phi_{\bf G'}(\kv,\rv). 
\ee

In conclusion, the core of the FLAPW self-consistent scheme is formed by a sequence of outer iterations, each one containing multiple large dense generalized eigenpencils. In order to numerically compute the charge density $\nr$ at each iteration, the matrices $A$ and $B$ need to be initialized for each $\kv$-point and the generalized eigenproblem $Ax = \lambda Bx$ solved. These two computing instances are the most machine-time consuming part of each iteration. It is our intention in this short paper to present some preliminary results and future ideas to reduce the computational time and improve the performance of the self-consistent cycle.
\vspace{-0.3cm}
\section{A new computational philosophy}
It is important to look at the entire DFT iterative process as a series of correlated problems $\left\{P_i\right\}$: it starts from an initial charge density $n_0({\bf r})$, executes a series of iterations $P_1 \dots P_i P_{i+1} \dots P_N$ and converges to a final density $n_f({\bf r})$. The problems are correlated since the output of a certain iteration $P_i$ is used as input for the next one $P_{i+1}$. Current codes that implement FLAPW (i.e. FLEUR, WIEN, FLAIR, Exciting, ELK) follow a simple approach: 1) compute every mathematical quantity involved in the solution of each $P_i$, 2) use libraries to solve the generalized eigenproblems as black boxes. This approach allows for standard and accurate calculations but is quite time-consuming and often sub-optimal. Our philosophy follows a counter-intuitive path: increase the number of iterations $\tilde{P}_i \;,\; i= 1 \dots M \;,\; M \gg N$ but make them fast so that the overall process $\left\{\tilde{P}_i\right\}$ is faster than $\left\{P_i\right\}$.

We plan to achieve the target by acting on the two most expensive sections of the self-consistent cycle, namely the matrix initialization and the eigenproblem solutions. Based on the observation that a considerable amount of matrix entries do not change between iterations, we want to save on computations that update the Hamiltonian $A$ and the overlap matrix $B$. At the same time we will show that eigenvectors from one iteration can be used to compute the eigenpairs of the next. Clearly, using information from previous iterations will speed up execution time of the single $\tilde{P}_i$. It will also make each $\tilde{P}_i$ less accurate, requiring a higher number of iterations in order to reach convergence. In the end we believe that the advantage gained by speeding up each single $\tilde{P}_i$ will overcome the disadvantage of having a larger number of iterations leading to a net saving for the set of correlated problems $\left\{\tilde{P}_i\right\}$.

We have conducted a preliminary study running a DFT process using the FLEUR code~\cite{FLEUR} to simulate two simple metallic systems: bulk copper (Cu\_b) and an iron multilayer (Fe\_l). The simplicity of the systems allows for a rather small matrix size M, where M$\simeq$50 for Cu\_b and M$\simeq$400 for Fe\_l. For the copper system, the simulation starts converging rapidly after the 4-5th iteration while the iron system starts converging after the 8-9th iteration. For each step of the two convergence processes we have collected the $A$ and $B$ matrices for each $\kv$-point and analyzed them using either a specifically designed MatLab routine or existing libraries.

\begin{figure}[h]\label{Eig}
	 \includegraphics[width=8cm,height=4.5cm]{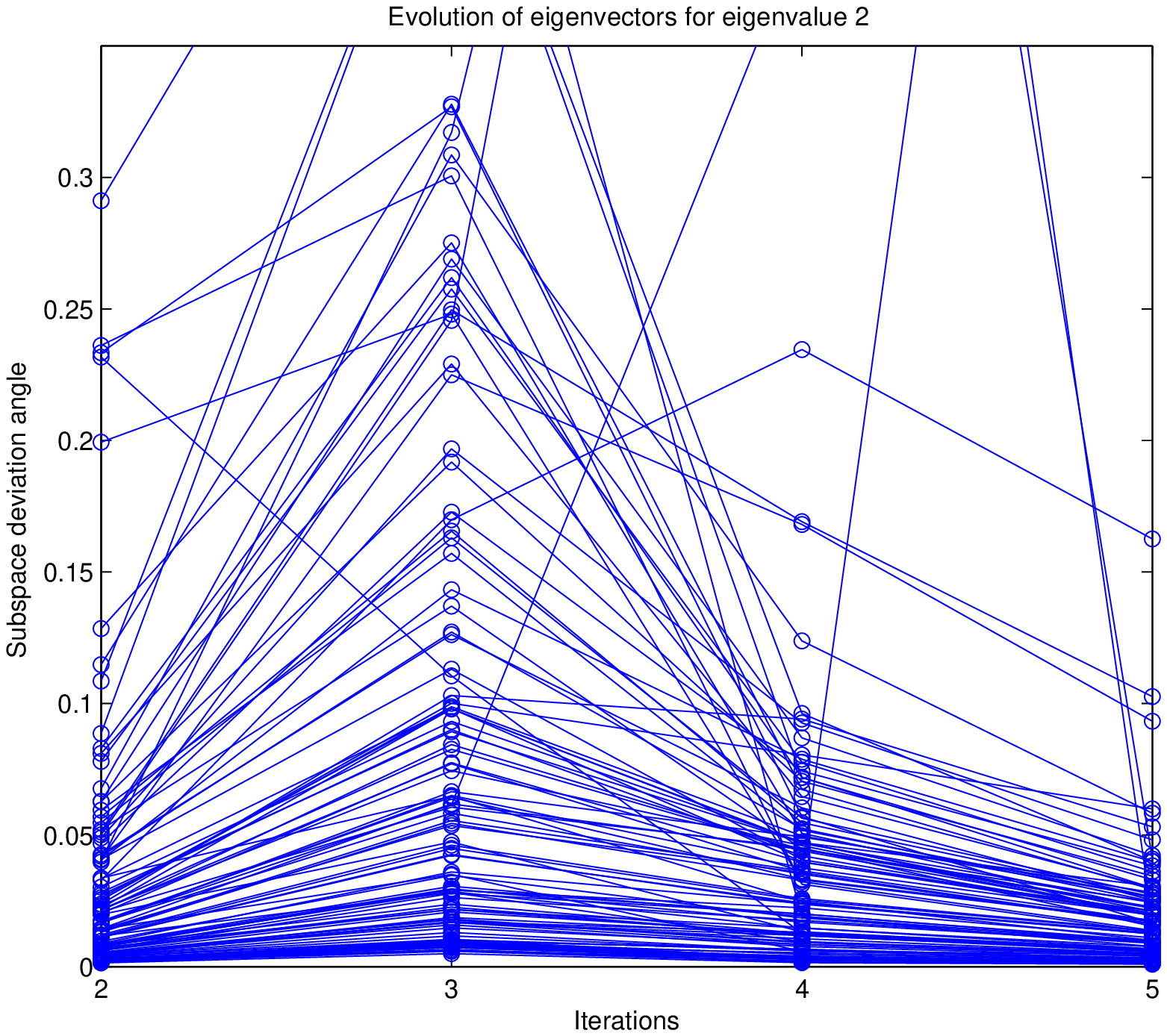}~~~~\includegraphics[width=8cm,height=4.5cm]{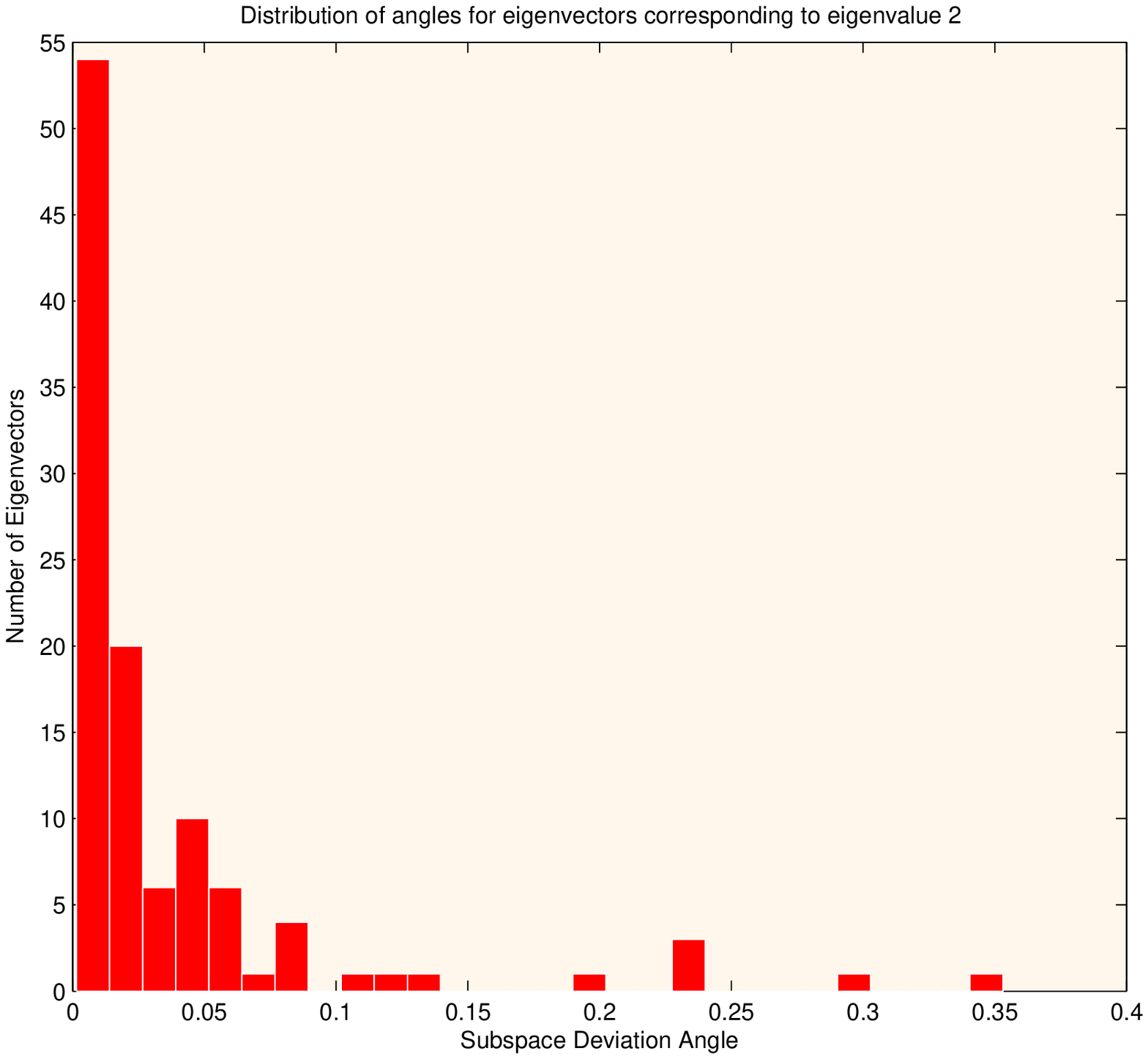}
\caption{Cu\_b $N=5$ case. On the left: the evolution of the subspace deviation angle for each of the $\kv$-point eigenvectors corresponding to the 2nd lowest eigenvalue. Angles exceeding graph borders correspond to degenerate eigenspaces. On the right: distribution of deviation angles for the same set of eigenvectors limited to the second iteration. All the angles are normalized to one (1.00 = $\frac{\pi}{2}$).}
\end{figure}   

\noindent
{\fontsize{11}{7}\selectfont \bf Eigenvectors evolution ---}
In order to study how eigenvectors evolve from one iteration to the next, we needed to establish a criterion that allows comparison of each eigenvector of $A$ at iteration $i$ with its corresponding eigenvector at the following iteration $i+1$. This is not a trivial task since the indexing of a set of eigenvalues, ordered by magnitude, can change substantially from one iteration to the next. The indexing of the eigenvectors meets the same fate, interfering with the ability to find a one-to-one correspondence between vectors of neighboring iterations. On the contrary, subspace deviation angles between corresponding eigenvectors of adjacent iterations should take comparable values and be uniformly distributed. Taking advantage of the latter observation, we designed an angle minimization procedure that made it possible to re-index the eigenvectors of the new iteration. The deviation angles were then computed from simple scalar products between re-ordered eigenvectors of adjacent iterations and were used as indicators of the evolution of the vectors.

We systematically looked at the deviation angles of a generic eigenvector $x(\kv,\lambda_i,n)$, corresponding to eigenvalue $\lambda_i$, as it evolves from iteration $n=1$ to iteration $n=N$. 
In all of the cases analyzed, the angles start decreasing monotonically after the first 2-3 iterations until they become almost negligible when the process $\left\{P_i\right\}$ is about to converge. Even at $n=2,3$ the average of the deviation angle among all the eigenvectors for all $\kv$-points is quite small $\vev{\theta_{dev}} = 0.04$ (1.00 = $\frac{\pi}{2}$) and its distribution quite peaked ($\sigma_{\theta_{dev}}$ = 0.135) (Fig.~\ref{Eig}).

Based on this study we have implemented a process based on our new computational philosophy. We start with few initial standard iterations $P_1\dots P_i \quad i \ll N$ computed with the full direct eigensolver so as to allow the simulation to bypass the phase with relatively large deviation angles. Then a series of fast iterations $\tilde{P}_i\dots \tilde{P}_{j-1}$ are carried out using the eigenvectors from prior iterations to initialize an iterative eigensolver. This procedure continues until progress towards convergence flattens out. At this point we re-start with a full data iteration $P_i$ in order to refresh the process, followed by another series of fast iterations $\tilde{P}_{j+1}\dots \tilde{P}_{k-1}$, and so on.

We have tested this procedure on both systems Cu\_b and Fe\_l using the LAPACK~\cite{LAPACK} QR algorithm for the eigenproblems of the $P$ iterations and the ARPACK~\cite{ARPACK} Implicit Restarted Arnoldi Method (IRAM) for eigenproblems of the fast iterations $\tilde{P}$. The results are somewhat encouraging but also show that the blind use of a Krylov process to implement the fast iterations does not work in all cases. In fact, while in the Fe\_l case the $\tilde{P}$s were more than five times faster than $P$s, the opposite was true for the Cu\_b case. These results suggest a more careful approach should be used in implementing the fast iterations.
\vspace{0.3cm}\\
\noindent
{\fontsize{11}{7}\selectfont \bf Matrix generation ---}
In a second set of experiments, we compared Hamiltonian matrices $A$ at the same $\kv$-point between adjacent iterations. The objective was to understand how much variation the matrix entries would undergo from one cycle to the next and how much this variation depends on the progress towards convergence of the entire DFT process. A favorable answer to this question would allow us to apply the new computational philosophy in the same form expressed for eigenvectors evolution. The difference would be that now the series of fast iterations $\tilde{P}_i\dots \tilde{P}_{j-1}$ would be represented by generalized eigenproblems involving matrices where only a fraction of the entries have been updated.

\begin{figure}[h]\label{Mat}
	 \includegraphics[width=7.5cm]{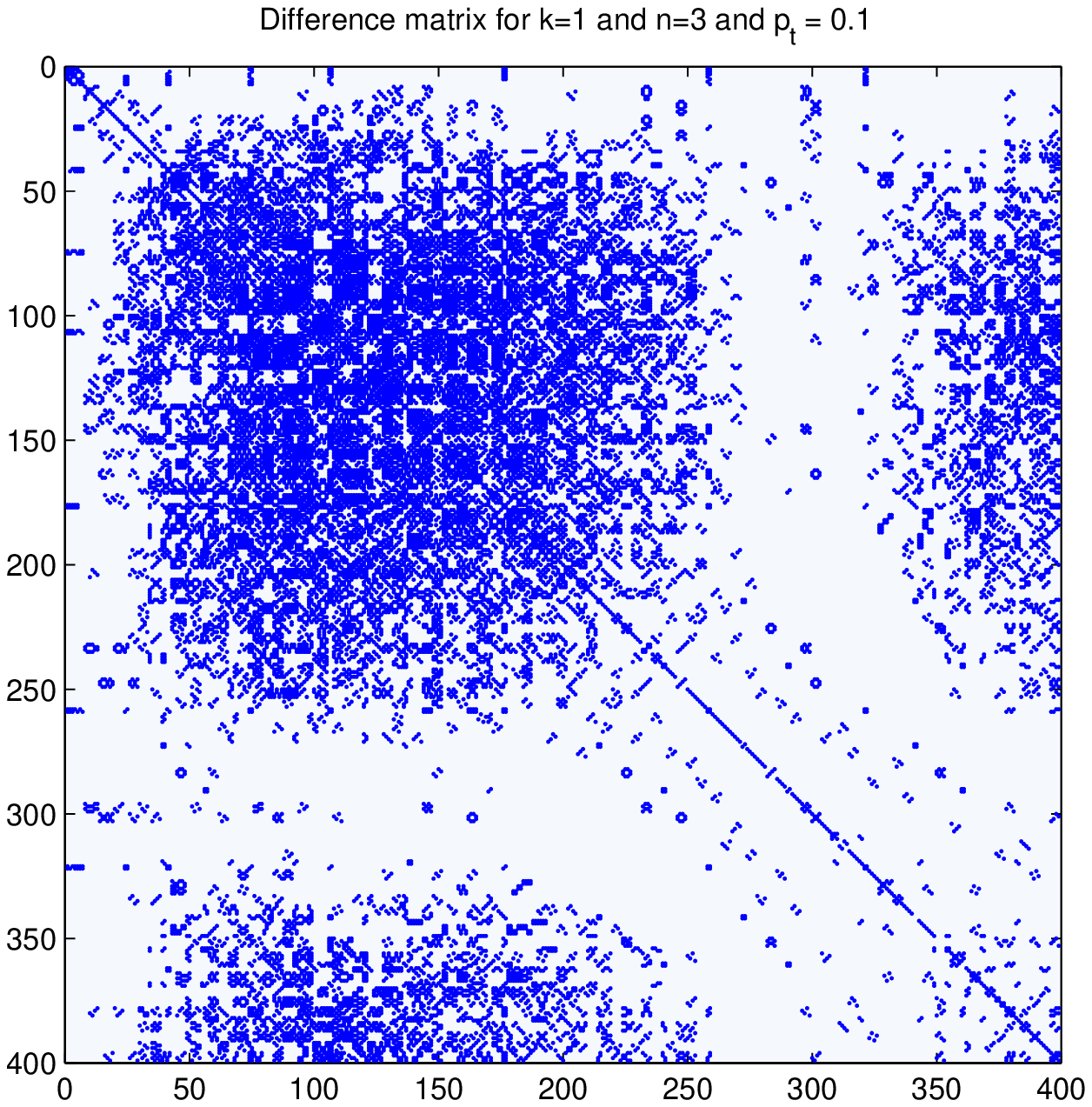}~\includegraphics[width=7.5cm]{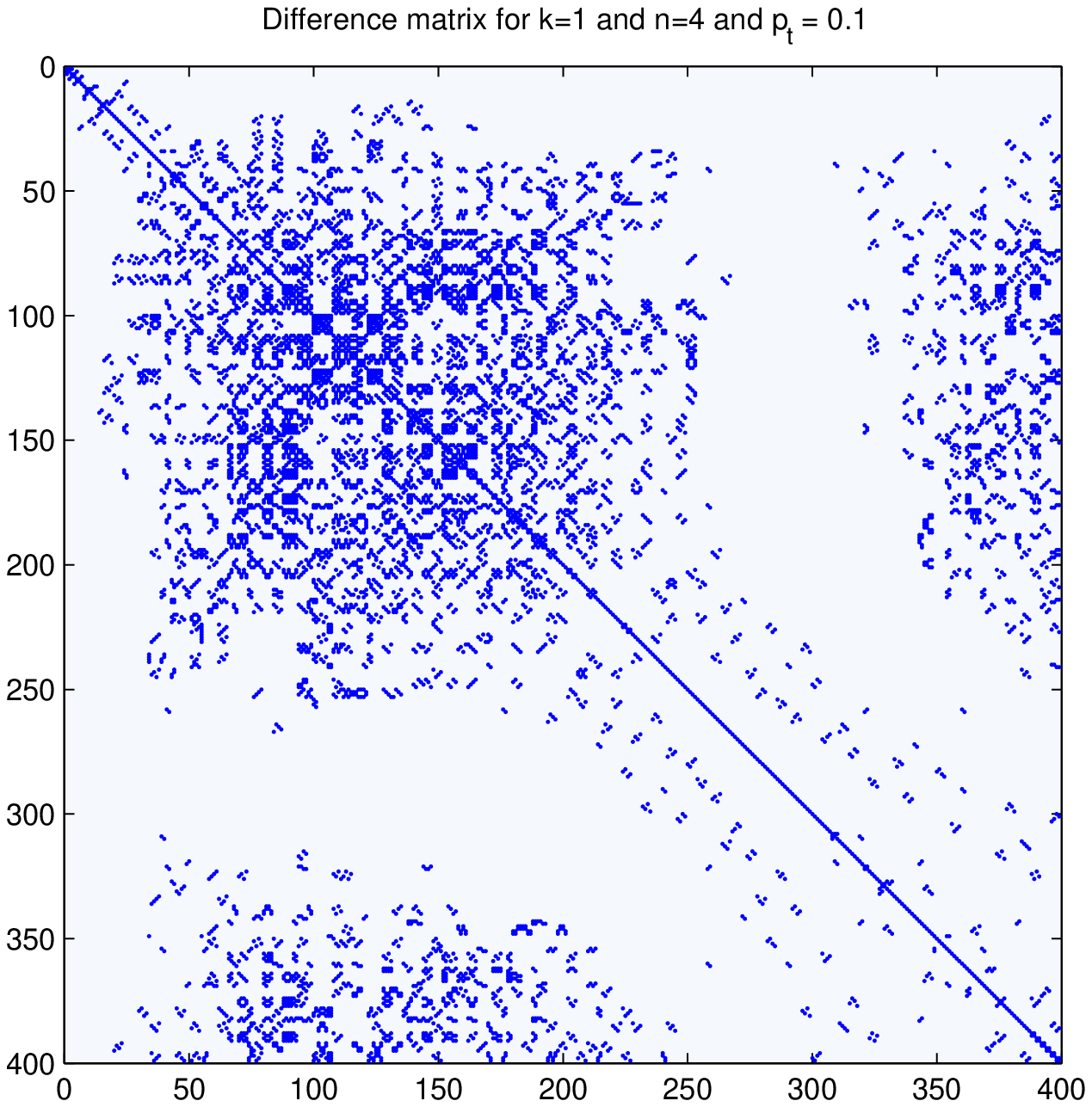}
\caption{Fe\_l $N=11$ case. On the left: plot of $\tilde{A}(\kv=1,n=3)$ for $p_t = 0.1$ On the right: plot of $\tilde{A}(\kv=1,n=4)$ for $p_t = 0.1$. Cross-like unchanging patterns are visible in both plots. At fixed $p_t$ the amount of entries above threshold has clearly decreased from $n=3$ to $n=4$.}
\end{figure}  

First we had to establish the most appropriate ``metric'' to gauge when a variation is substantial and when it is not. In the end we chose as a metric the maximal entry variation for each matrix difference $\delta_{n+1} = \max(A(\kv,n+1) - A(\kv,n))$  and normalized the entries of the difference with respect to it. The resulting matrices $\tilde{A}(\kv,n+1) = \frac{A(\kv,n+1) - A(\kv,n)}{\delta_{n+1}}$ were analyzed for different values of the variation threshold and for different values of the iteration level $n$. For the sake of conciseness, we present here some sample results taken only from the Fe\_l system. Similar results have been obtained for the Cu\_b and other simple crystals.

Our qualitative study shows that no matter how low the threshold value $p_t$ (measured in percentage of $\delta_{n+1}$) is, the set of difference matrices $\{\tilde{A}(\kv,n)\}$ shows extended areas where entries do not vary considerably. Moreover, as we progress to higher iteration levels $n \rightarrow N$, the number of entries of $\tilde{A}(\kv,n)$ that undergo a variation above threshold decreases monotonically (Fig.~\ref{Mat}). This behavior is expected since, as we advance towards convergence, the set of basis functions $\phi_{\bf G}(\kv,\rv)$ needs lesser refinements in order to minimize the ground energy of the quantum mechanical system. In conclusion, it seems that the Hamiltonians (we expect a similar behavior from the overlap matrices) follow a universal behavior independent from the physical system under consideration. This fact implies that we can save a great deal of computational time if we avoid updating the corresponding matrices. For instance if, in the Fe\_l case, we refrained from updating the entries of $A(\kv,4)$ for a threshold value $p_t = 0.1$, we would save computing about 40\% of the entries - quite a substantial portion of the total time spent generating the entire matrix. 

\vspace{-0.3cm}
\section{Summary and conclusions}
In this short paper we illustrated a new computational approach to DFT-based methods. Two promising research directions emerge from our preliminary study: exploit limited eigenvectors evolution and save on matrix entry generation. Several steps still require investigation in this line of research. Current work on eigenvectors evolution is focusing on building a set of tailored algorithms that would efficiently use the information provided by eigenvectors of prior iterations. The target is to achieve consistency even for physical systems of small dimensions or special geometry. Major candidates, in this line of research, are a new eigensolver based on the Sakurai-Sugiura~\cite{SS} method and an improved version of the subspace iteration method. In matrix generation we are working to clarify the relation among threshold value $p_t$, speed up of the single iteration $\tilde{P}$, and the accuracy of solutions of the generalized eigenvalue problems. It is crucial that the fast iterations not loose too much accuracy in their advance towards convergence while gaining in speed. A clear ratio between speed up and progress of the entire process needs to be set and correlated with the choice of threshold value for each iteration. In a following phase we want to develop an automated self-tuning slider that adjusts the value of $p_t$ in order to optimize the trade-off between speed and accuracy. 

Last but not least, it is evident that there are clear patterns in the unchanging portions of the matrices. This fact strongly points to a more profound explanation that is likely connected to the physics embedded in the self-consistent cycle. If we can uncover the physical rationale behind the unchanging patterns, we can directly influence the choice of orbital functions and improve the DFT method before the numerical computation is initiated. This would constitute a very important contribution to any DFT-based method reversing the usual flow of information that goes from theory to computation. It would establish a unique example where computer simulations provide insight into theoretical models by simple analysis of the numerical inputs and outputs. 

\vspace{-0.3cm}
\begin{theacknowledgments}
Funding for this project has been provided by the J\"ulich Aachen Research Alliance (JARA-SIM) consortium through the JARA Seed Fund initiative. The authors would like to thank Prof.~S.~Bl\"ugel and Dr.~D.~Wortmann for useful discussions and collaborative efforts. Special thanks to M.~Petschow and D.~Fabregat for suggestions and support.
\end{theacknowledgments}
\bibliographystyle{aipproc}

\end{document}